\documentstyle[multicol,epsf,epsfig,aps,amssymb]{revtex}

\begin{document}
\draft
\title{Coulomb displacement energies, energy differences and neutron skins.}
\author{ A. P. ~Zuker$^a$, J. Duflo$^b$, S. M. ~Lenzi$^c$,
  G.~Mart\'{\i}nez-Pinedo$^{d,e}$, 
  A.~Poves$^f$, J.~S\'anchez-Solano$^f$}
\address{
(a) IReS, B\^at27, IN2P3-CNRS/Universit\'e Louis
Pasteur BP 28, F-67037 Strasbourg Cedex 2, France\\
(b) Centre de Spectrom\'etrie Nucl\'eaire et de Spectrom\'etrie
de Masse (IN2P3-CNRS) 91405 Orsay Campus, France\\
(c) Dipartimento di Fisica and INFN, Via F. Marzolo 8, I-35131 
 Padova, Italy\\
(d) Institut for Fysik og Astronomi, {\AA}rhus Universitet, DK-8000
{\AA}rhus C, Denmark\\
(e) Department f\"ur Physik und Astronomie, Universit\"at Basel, CH-4056
Basel, Switzerland\\
(f) Departamento de Fisica Te\'orica C-XI Universidad Aut\'onoma
de Madrid, E-28049, Madrid, Spain}
\date{\today}
\maketitle
\begin{abstract}
  A Fock space representation of the monopole part of the Coulomb
  potential is presented. Quantum effects show through a small orbital
  term in $l(l+1)$. Once it is averaged out, the classical
  electrostatic energy emerges as an essentially exact expression,
  which makes it possible to eliminate the Nolen-Schiffer anomaly, and
  to estimate neutron skins and the evolution of radii along yrast
  states of mirror nuclei. The energy differences of the latter are
  quantitatively reproduced by the monopole term and a schematic
  multipole one.
\end{abstract}
\pacs{PACS numbers: 21.10.Sf, 21.10.Ft, 21.10.Gv, 21.60.Cs, 27.40.+z}
\begin{multicols}{2}
  The electrostatic energy of a sphere of radius $R$ and charge $Ze$
  is easily calculated to be $E_{Cs}=3e^2Z^2/5R$ ($E_{Cs}$
  stands for ``simple Coulomb energy''). It is under this guise that
  the Coulomb field enters the Bethe-Weiz\"acker mass formula, and
  becomes a basic quantity in nuclear structure. Direct evidence of
  entirely Coulomb effects has long been available from displacement
  energies between mirror (or in general, analog) ground states
  (CDE)~\cite{BM69,NS69}, and more recently from differences in yrast 
    excitation energies  in mirror $pf$-shell nuclei
  (CED)~\cite{cameron,49,47-49,51,50}.
  
  The CDE  range from few to tens of MeV. They should be given mainly
  by $E_{Cs}$, but are underestimated by mean-field calculations: the
  Nolen-Schiffer (NS) anomaly~\cite{NS69}, still an open problem.
  
  The CED  are very small (of the order of 10-100 keV). They 
  have been semiquantitatively explained by shell model calculations
  using Coulomb matrix elements but there is some room for
  improvement.
  
  The present letter intends to give a unified microscopic description
  of CDE and CED, by separating the Coulomb field into monopole and
  multipole components $V_C=e^2(M\omega/\hbar)^{1/2}/r=V_{Cm}+V_{CM}$
  following ref.~\cite{dz96}(DZ). The monopole $V_{Cm}$ contains all
  terms quadratic in scalar products of Fermion operators
  $a_i^{\dagger}\cdot a_j$. Its diagonal part involves only proton
  number operators, and should be responsible for $E_{Cs}$, and hence
  for the observed CDE.  The non diagonal part will not be considered
  here: it leads to isospin mixing, but energetically its effect is
  small. The multipole $V_{CM}$ contains all non-monopole matrix
  elements and accounts for much of the CED.
  
  As an introduction we examine the origin of the NS
  anomaly, by calculating $E_{Cs}$ with proton radii of the form
  \begin{equation}
    \label{eq:R}
R_{\pi}=r_0\, A^{1/3}\left(1-\upsilon(\frac{t}{A})^2-\zeta
\frac{t}{A^{4/3}}\right)e^{(1.75/A)}+{\cal D}.     
\end{equation}
($A=N+Z$, $t=N-Z$, $\pi,\, \nu\equiv$ protons, neutrons.) 

Isospin conservation is assumed, which implies that $R_{\nu}$ is the
same as $R_{\pi}$ {\em for the mirror nucleus}, obtained by
interchanging $N$ and $Z$.  Therefore, in Eq.~(\ref{eq:R}), for $N>Z$,
$\upsilon>0$ represents a uniform contraction of the two fluids, while
$\zeta>0$ represents a $\pi$-contraction and a $\nu$-dilation. Hence, a
(fractional) neutron skin can be defined as
\begin{equation}
  \label{eq:skin}
  \Delta(\zeta)=\frac{R_{\nu}-R_{\pi}}{r_0\, A^{1/3}\,
  e^{1.75/A}}=\frac{2|t|\zeta}{A^{4/3}}, \quad N>Z. 
\end{equation}

The exponential factor takes care of the increase in $R_{\pi}$
observed in the light nuclei.  

${\cal D}$ is a phenomenological term that accounts for shell and
deformation effects. It is a sum of two quartic forms $\lambda
S_{\pi}S_{\nu}+\mu Q_{\pi}Q_{\nu}$ that vanish at the spin-orbit (or EI:
extruder-intruder) closures at $N$ or $Z=6,\, 14,\, 28,\, 50,\, 82,\,
126$. Defining $D_{\pi}=(p_{\pi}+1)(p_{\pi}+2)+2$, the degeneracy of
the EI$_{\pi}$ shell (e. g., $p_{\pi}=3$ for  $Z=28$ to 50),
$D_{r\pi}=p_{\pi}(p_{\pi}+1)$, the degeneracy of the non-intruder
subshells; the factors are $S_{\pi}=z(D_{\pi}-z)/D_{\pi}^2$ and
$Q_{\pi}=z(D_{r\pi}-z)/D_{\pi}^2$ ($z$= number of valence protons).
The parametrization is a variant of the
ones in~\cite{d94,dz99}.

Fits to $R_{\pi}=\sqrt{5\langle r_{\pi}^2\rangle/3}$---where $\langle
r_{\pi}^2\rangle$ is the measured mean square radius---for 634 nuclei
(with $N\ge Z$ except two cases!)  yield ($r_0$ and rmsd in fm)

\noindent
$r_0$=1.236 \ \ $\upsilon$=0.00\ \ $\zeta$=0.94\ \ $\lambda$=6.2
\ \ $\mu$=14.6 \ \ rmsd=0.018:

\noindent
A good fit with a Huge Skin (HS hereafter).
 
\noindent
$r_0$=1.220 \ \ $\upsilon$=0.61\ \ $\zeta$=0.00\ \ $\lambda$=5.7
\ \ $\mu$=27.0 \ \ rmsd=0.012:

\noindent
A much better fit with Zero Skin (ZS hereafter).
 
\noindent
$r_0$=1.226 \ \ $\upsilon$=0.45\ \ $\zeta$=0.29\ \ $\lambda$=5.7
\ \ $\mu$=24.0\ \  rmsd=0.011:

\noindent
An even better fit with a Minute Skin (MS hereafter).
 
Adding an exchange term to $E_{Cs}$, replacing $Z^2$ by $Z(Z-1)$
(conceptually better, as $V_C$ is two-body), the main contribution to
the monopole energy takes the form
  \begin{equation}
    \label{eq:vc}
  E_{Cm}=\frac{d\, Z(Z-1)(1-c\,Z^{-h})}{\rho},\quad
d=\frac{3e^2}{5r_0}=\frac{0.864}{r_0},
\end{equation}
$d$ in MeV and $\rho=(R_{\pi}-{\cal D})/{r_0}$: The $\cal D$
correction can be left out for simplicity as it does not affect what
follows; the high quality fits become average ones (rmsd trebled), but
still sufficient for our purpose.  For the exchange term the usual
choice is $h=2/3$, while $c$ varies appreciably~\cite{NS69}. Here we
set $h=1$ (explained in and after Eq.~(\ref{eq:vcapp})). The overall
factor $d$ is fixed. {\em Nevertheless}, it will be allowed to vary,
under the name $d_f$, together with $c$, in a fit to the 183 available
CDE from reference~\cite{aw95}. The $\chi^2$ values are calculated by
adding 200 keV to the experimental errors, to account for
uncertainties in the calculations. Obviously, consistency between
$R_{\pi}$ and $E_{Cm}$ demands $d_f=d$. The three fits to $R_{\pi}$
lead to:

\vspace{3pt}

\noindent
$d$=0.699\ \ $d_f$={\bf 0.77}\ \ $c=-0.5$\ \ $\upsilon$=0.00\ \
$\zeta$=0.94\ $\chi^2$=0.91:

\noindent
HS leads to a large overestimate of $d$. In other words:
keeping the correct $d$ leads to a large {\em underestimate} of the
CDE: the NS anomaly~\cite{NS69,a83}, unresolved to this day (see
however~\cite{fayans}).

\vspace{3pt}

\noindent
$d$=0.708\ \ $d_f$=0.69\ \ $\, c$=\ 1.3\ \ $\upsilon$=0.61\ \ 
$\zeta$=0.00\ $\chi^2$=1.46:

\noindent
ZS leads to a small underestimate of $d$.

\vspace{3pt}

\noindent
$d$=0.705\ \ $d_f$=0.71\ \ $\; c$=\ 0.9\ \ $\, \upsilon$=0.45\ \ 
$\zeta$=0.29\ $\chi^2$=1.20:

\noindent
MS leads to $d_f\approx d$. {\em The NS anomaly disappears}.
Clearly, $\zeta\lessapprox 0.3$ is a good guess in
Eq.~(\ref{eq:skin}).

\vspace{3pt}

The NS anomaly occurs because mean field calculations {\em that yield
  good} $R_{\pi}$ systematically overestimate $R_{\nu}$:
Ref.\cite{bartel} contains a nice illustration of the problem.  It is
ironical to note that a small neutron skin was recognized as a
possible solution of the anomaly---but rejected---by Nolen and
Schiffer~\cite{NS69}. Many reasons explain why this rejection held for
so long. We retain only two: neglect of the $\upsilon (t/A)^2$ term in
Eq.~(\ref{eq:R}), and lack of confidence in the validity of $E_{Cm}$.
As we show next, $E_{Cm}$ in Eq.~(\ref{eq:vc}) must be trusted, as
it is basically an exact form of $V_{Cm}$.

\vspace{3pt}

\begin{figure}[h]
    \begin{center}
      \leavevmode
      \epsfig{file=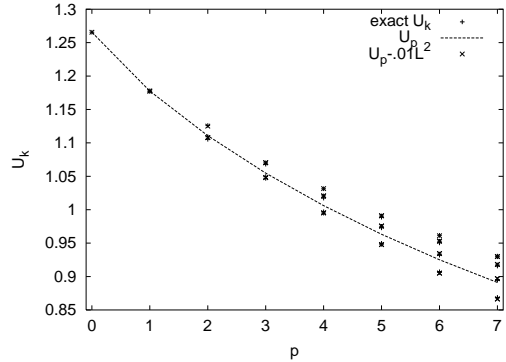,width=7cm}
   \caption{Dominant factor in the DZ decomposition of $V_{Cm}$}
      \label{fig:dz}
    \end{center}
\end{figure}

By definition, the diagonal monopole part of $V_C$ is ($[J]=2J+1$):
\begin{equation}
\label{eq:mono}
V_{Cm}=\sum_{i\leq k}\frac{z_i(z_k-\delta_{ik})}{1+\delta_{ik}}V_{ik}, \quad 
V_{ik}=\frac {\sum_{J}V_{Cikik}^J[J]}{\sum_{J}[J]}.
\end{equation}
The label $k\equiv plj$ stands for the quantum numbers specifying a
given harmonic oscillator (ho) orbit ($p$ is the principal quantum
number).  Restricting the sum to the first $\kappa$ major shells
containing $\tau$ orbits, $V_{Cm}$ is brought to a sum of factorable
terms by diagonalizing the matrix $\frac{1}{2}\{V_{ik}\}$ through the
unitary transformation $\cal U$:
\begin{equation}
  \label{eq:vcfac}
  V_{Cm}=e^2\sqrt{\frac{M\omega}{\hbar}}\sum_n \left[{\cal E}_n\left(\sum_k
  z_k {\cal U}_{kn}\right)^2-{\cal C}\right],
\end{equation}
where ${\cal C}=\sum_n z_n\, V_{nn}/2$ is the one-body counterterm in
Eq.~(\ref{eq:mono}) left out of the diagonalization. By rescaling
${\cal E}_n=\tau E_n$, ${\cal U}_{kn}=(\tau)^{-1/2}U_{kn}$, the results
become independent of $\tau$.  To fix ideas choose $\kappa=8$, i. e.,
$\tau=36$. We expect to extract something close to the $Z^2$ {\em
  operator}, in which case only the highest eigenvalue ($E_{36}$)
should be non vanishing, with $U_{k\,36}=1$. The diagonalization
produces indeed an $E_{36}=0.192$ that is 30 times larger than the
next and over 100 times larger than the second next. (Increasing the
number of shells would only increase the number of negligible terms:
Eq~(\ref{eq:vcfac}) is an {\em exact} representation in Fock space.)
Fig.~\ref{fig:dz} shows that ($U_k\equiv U_{k\,36}$ ) is very well
approximated by the form $U_p-0.01\, L^2$,
\noindent
where $U_p=\sum_{j}(2j+1)U_k$ is the average of $U_k$ over $j$-values,
and $L^2=l(l+1)-p(p+3)/2$ is an $l(l+1)$ term referred to its
centroid. This result is still {\em almost exact}. The presence of
$L^2$ is interesting, but here we average it out, to transform the
{\em operator} $\sum_p z_p\, U_p$ into a {\em c-number} by taking
expectation values in Eq.~(\ref{eq:vcfac}) over ho closed shells.
Introducing $b=(\hbar/M\omega)^{1/2}$, we find
\begin{eqnarray}
  \label{eq:vcapp}
  \langle V_{Cm}\rangle&\approx&\frac{e^2}{b}
 \left[ E_{36}\langle\sum_p z_p U_p\rangle^2-\langle {\cal
 C}\rangle\right] \\
   &\approx&\frac{e^2}{b}\left[
  E_{36}\left(\sum_p  (p+1)(p+2)U_p\right)^2-\langle {\cal
  C}\rangle\right]
\nonumber\\
&\approx&\frac{e^2}{b} 0.445\,
  (Z(Z-1))^{(1-1/12)}(1-0.96/Z).\nonumber 
\end{eqnarray}
The last line is a numerical fit to the previous one.
$\langle\sum_p z_p U_p\rangle\approx 1.522Z^{1-1/12}(1-0.48/Z)$, so
$0.445=0.192\times (1.522)^2$). In the exchange term: $Z^{-1}$ is
better than $Z^{-2/3}$; 0.96 is close to $c$ in ZS and MS CDE fits.
The apparently awkward 1-1/12 exponent comes out of the fit It will be
seen to be natural once we extract $b$ from $\hbar\omega_{\pi}$
calculated as in \cite[Eq.~(2-157)]{BM69}, but treating separately
neutrons and protons:
\begin{equation}
  \label{eq:hwz}
  \hbar\omega_{\pi}=\frac{35.59(2Z)^{1/3}}{\langle r_{\pi}^2\rangle}\,
  \Longrightarrow\, \frac{e^2}{b}\, 0.445=
  \frac{0.860Z^{1/6}}{r_0\, \rho},  
\end{equation}
where we have used $\langle r_{\pi}^2\rangle=3r_0^2\rho^2/5$ to bring
Eq.~(\ref{eq:vcapp}) to the form of Eq.~(\ref{eq:vc}). The factor in
$Z^{1/6}=(Z^2)^{1/12}\approx (Z(Z-1))^{1/12}$ in $(\hbar\omega)^{1/2}$
conveniently corrects the ``awkward exponent'' and both equations
become identical to within a 0.5\% discrepancy in the $d$ coefficient:
$0.860/r_0$ in Eq.~(\ref{eq:vcapp}) {\it vs.}  $0.864/r_0$ in
Eq.~(\ref{eq:vc}). Therefore, $E_{Cm}$ is essentially exact to within
the averaging of $L^2$. A full treatment of this term is of obvious
interest.

Let us turn to the CED in the $pf$ shell. They are differences in
expectation values of $V_C$ for excitation energies;
CED$_J=E_C^x(Z,N,J)-E_C^x(N,Z,J)$ ($Z>N$). The wavefunctions are
obtained through standard shell model
calculations~\cite{ANTOINE,47-49,51,50} (they depend {\em very} little
on the interaction: KB3, KB3G or FPD6). First we separate monopole and
multipole pieces:
\begin{equation}
  \label{eq:cedj0}
  {\rm CED}_J=\Delta\langle V_{Cm}\rangle_J+\Delta\langle
  V_{CM}\rangle_J
\end{equation}
$\langle V_{Cm}\rangle_J$ is proportional to the difference of
(inverse) radii between a $J$ yrast and the ground state:
$R_J^{-1}-R_0^{-1}\approx (R_J-R_0)/R_0^2)$. Since $R_J$ is very
nearly the same for both members of the mirror pair (remember: the
neutron skin is small), it will be proportional to the {\em average
  neutron plus proton} occupancies for the individual orbits, which we
denote by $\langle m_{k}\rangle_J/2$, with $m_k=z_k+n_k$ ($n_k$ is the
number of neutrons). Now: to good approximation, the $R_0^2$
denominator is a constant over the region of interest ($A=47$-51) ,
furthermore, it is reasonable to assume that radii of the $pf$ orbits
depend only on $l$, and, finally, the $p_{1/2}$ occupancy is always
negligible. Therefore, $\langle V_{Cm}\rangle_J$ can be taken to
depend only on $\langle m_{p_{3/2}}\rangle_J/2$.

The multipole contribution $\Delta\langle V_{CM}\rangle_J$ is given by
the expectation values of the effective Coulomb potential in the $pf$
shell. Then
\begin{equation}
  \label{eq:cedj1}
  {\rm CED}_J=a_m\langle m_{p_{3/2}}\rangle_J+\Delta\langle
  V_{Cpf}^{eff}\rangle_J 
\end{equation}
The value of $a_m$ can be estimated by noting that the single particle
$p_{3/2}$ state in $^{41}$Sc is 200 keV below its analogue in
$^{41}$Ca. This number comes from two effects: a larger radius that
depresses the $p_{3/2}$ orbit, and the single particle $L^2$ term in
Fig.~\ref{fig:dz} that depresses the $f_{7/2}$ ground state orbit. The
latter effect is readily found to be $\approx$ 150 keV by expanding
$(\sum_k U_p+.01L^2)^2$ around the $A=40$ closed shell and using the
numbers in, and after Eq.~(\ref{eq:vcapp}). Then, $a_m\approx
(.200+.150)/2=0.175$ MeV. Note that the single particle contribution
in $L^2$ is proportional to the {\em difference} of proton and neutron
occupancies. It is important in $A=41$, but typically ten times
smaller than the radial effect in $A=$47-51, so we have neglected it.

The available information on $\Delta\langle V_{Cpf}^{eff}\rangle_J$
involves only the $f_{7/2}$ matrix elements extracted from the
$^{42}$Ti-$^{42}$Ca pair, which yields ($7\equiv f_{7/2}$) $V_{CM\,
  7777}^J\equiv V_{Cf_{7/2}}^{eff}=$ (86.9, 116.9, 10.9, -59.1) keV,
for $J$= 0, 2, 4, 6 respectively, to be compared with the ho values
$V_{Cf_{7/2}}^{ho}$= (81.6, 24.6, -6.4, -11.4) keV.  Since we are
dealing with $V_{CM}$, the centroids $V_{77}$ (Eq.~(\ref{eq:mono}))
have been removed. They are very close (304 keV for $A$=42, 308 keV
for ho) {\em because} $E_{Cm}$ {\em depends on conserved quantities
  that cannot be renormalized}. The multipole matrix elements are very
different, and the data unequivocally prefer the $A=42$
set~\cite{47-49,51}. The strategy adopted in these references was to
use a $V_{Cpf}^{eff}$ with $V_{Cf_{7/2}}^{eff}$, keeping
$V_{Cpf}^{ho}$ for the other matrix elements, which turned out to be
almost irrelevant: the $V_{Cf_{7/2}}^{eff}$ set by itself accounts for
the full $V_{Cpf}^{eff}$ chosen in this way.  The results alternated
between agreement and distorsion of the observed patterns.

As there is no justification in accepting an enormous renormalization
for the $f_{7/2}$ orbits and leave the rest of the interaction
unchanged, we shall attempt a more general treatment, by exploring the
possibility of writing an effective interaction solely in terms of
$V_{Cf_{7/2}}^{eff}$---properly renormalized to give a plausible
account of the full $V_{Cpf}^{eff}$.  First we check that the program
can be enacted for the ho set. We try
\begin{equation}
  \label{eq:horen}
\Delta\langle V_{Cpf}^{ho}\rangle_J=b\, \Delta\langle
V_{Cf_{7/2}}^{ho}\rangle_J+a\, \langle m_{p_{3/2}}\rangle_J:
\end{equation}
\begin{figure}[h]
    \begin{center}
      \leavevmode \epsfig{file=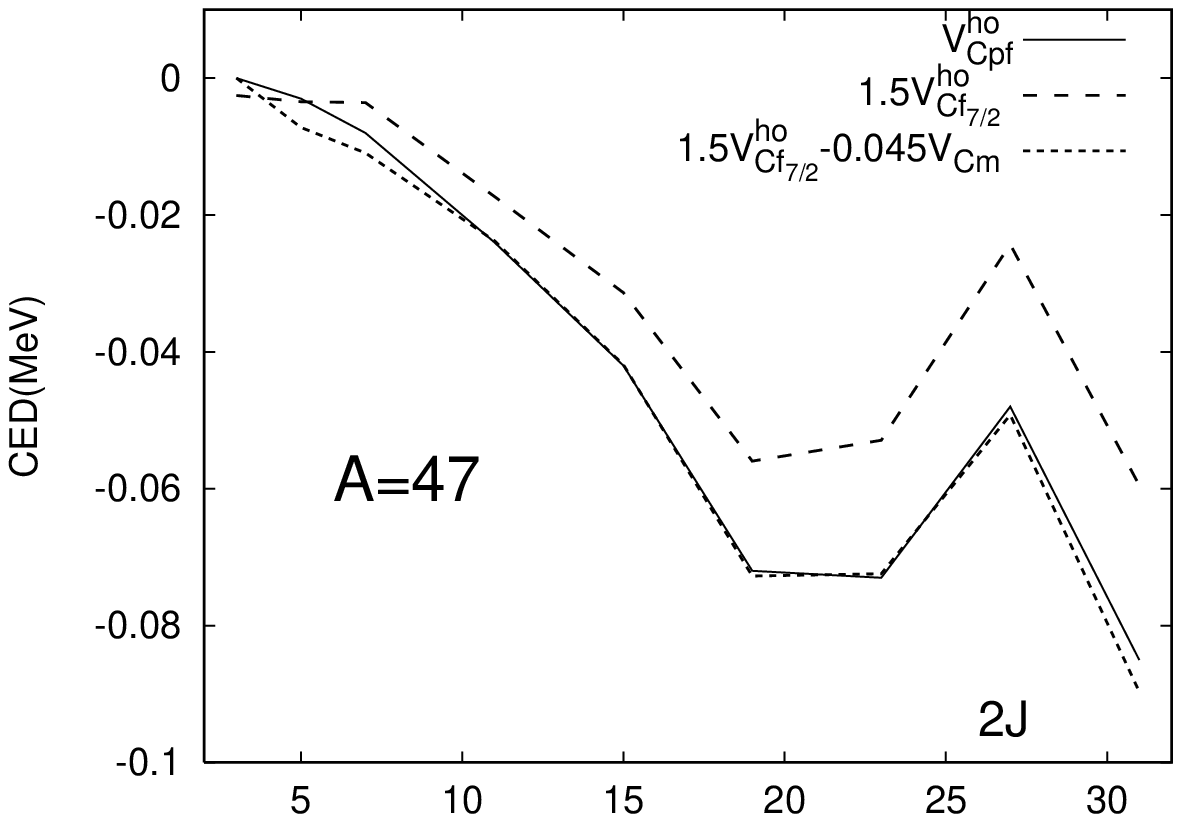,width=7cm}
      \epsfig{file=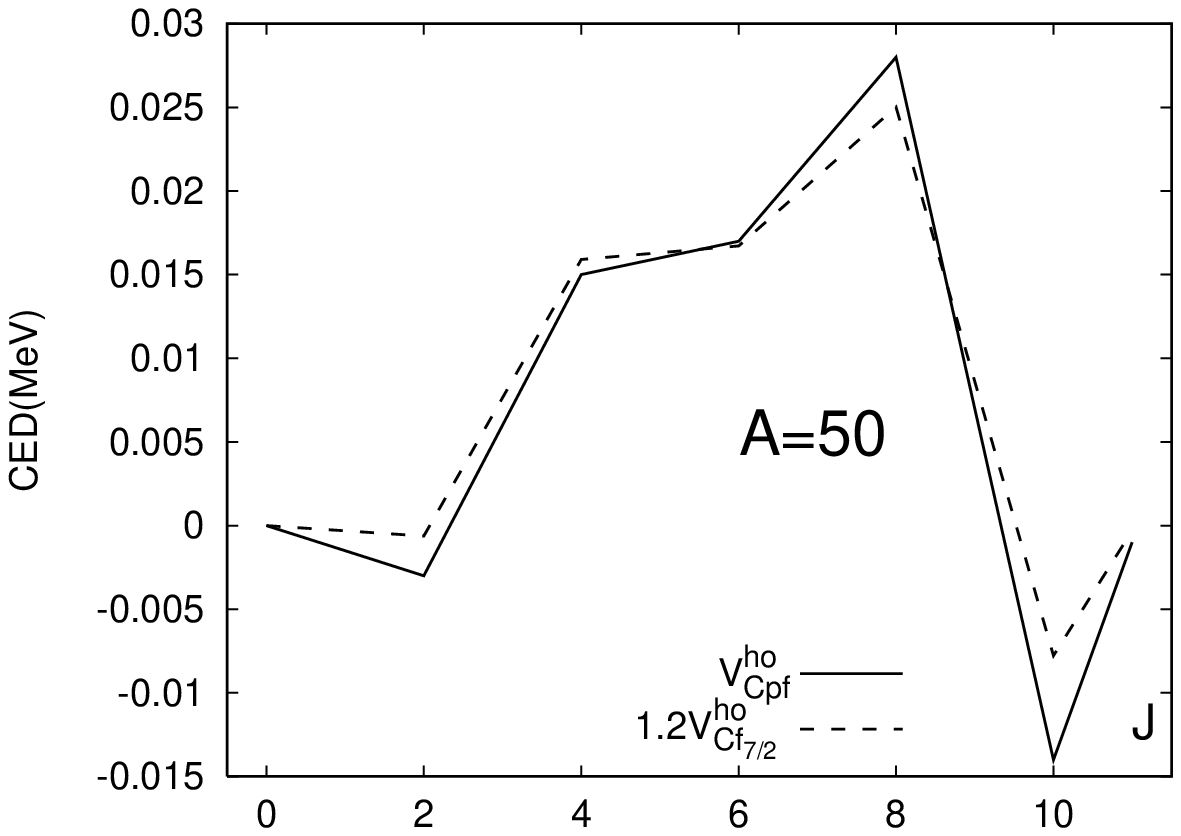,width=7cm}
   \caption{Calculated CED normalizations}
      \label{fig:tuttir}
    \end{center}
\end{figure}
Eq.~(\ref{eq:horen}) is only a {\em numerical recipe}, but it works
well, as seen in Fig.~\ref{fig:tuttir}, representative of the quality
of the adjustment in the four cases (parameters in
Tab~\ref{tab:param}).  Next, assume that the prescription applies to
the renormalized case and try to {\em invent} an effective interaction
consistent with the data and {\em respecting the condition that the $a_m$
parameter in Eq.~(\ref{eq:cedj1}) be a constant that should be
determined carefully}: There is no room for invention here.
\begin{table}[h]
\caption{Adopted parameters in Eq.~(\ref{eq:horen}) and Eq.~(\ref{eq:cedj2})} 
    \begin{center}
      \begin{tabular}{cccclcc}
$A$& $b$ &   $a$  &&  $b_M$     &$a_M$  &  $a_m$        \\
\hline                                                   
47 & 1.5 &-0.045  && 0.75       &-0.080 & 0.150         \\
49 & 1.5 &\ 0.000 && 0.75       &\ 0.000& 0.150         \\
50 & 1.2 &\ 0.000 && 0.60       &\ 0.000& 0.150         \\
51 & 1.6 &-0.030  && 0.80       &-0.054 & 0.150         
\end{tabular}
\label{tab:param}
\end{center}
\end{table}

\vspace{-.75cm}
  \begin{figure}[h]
    \begin{center}
      \leavevmode
      \epsfig{file=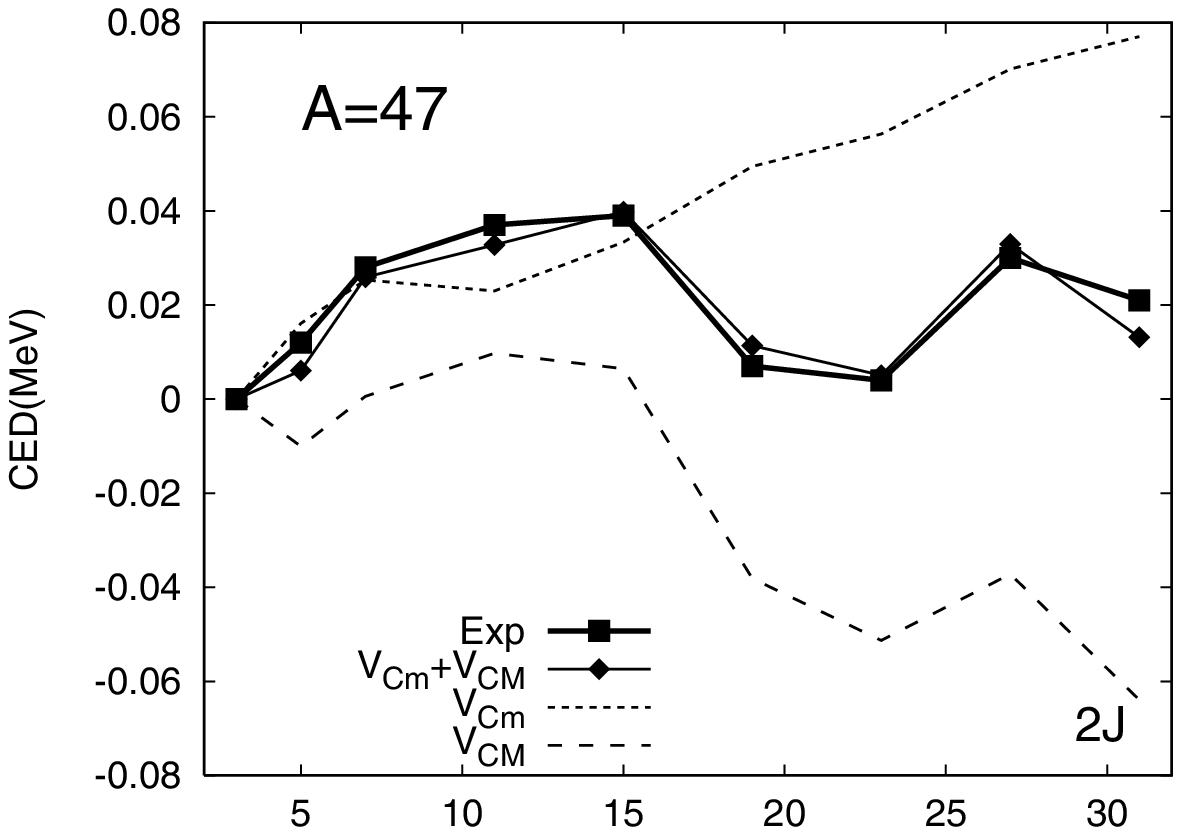,width=7cm}
      \epsfig{file=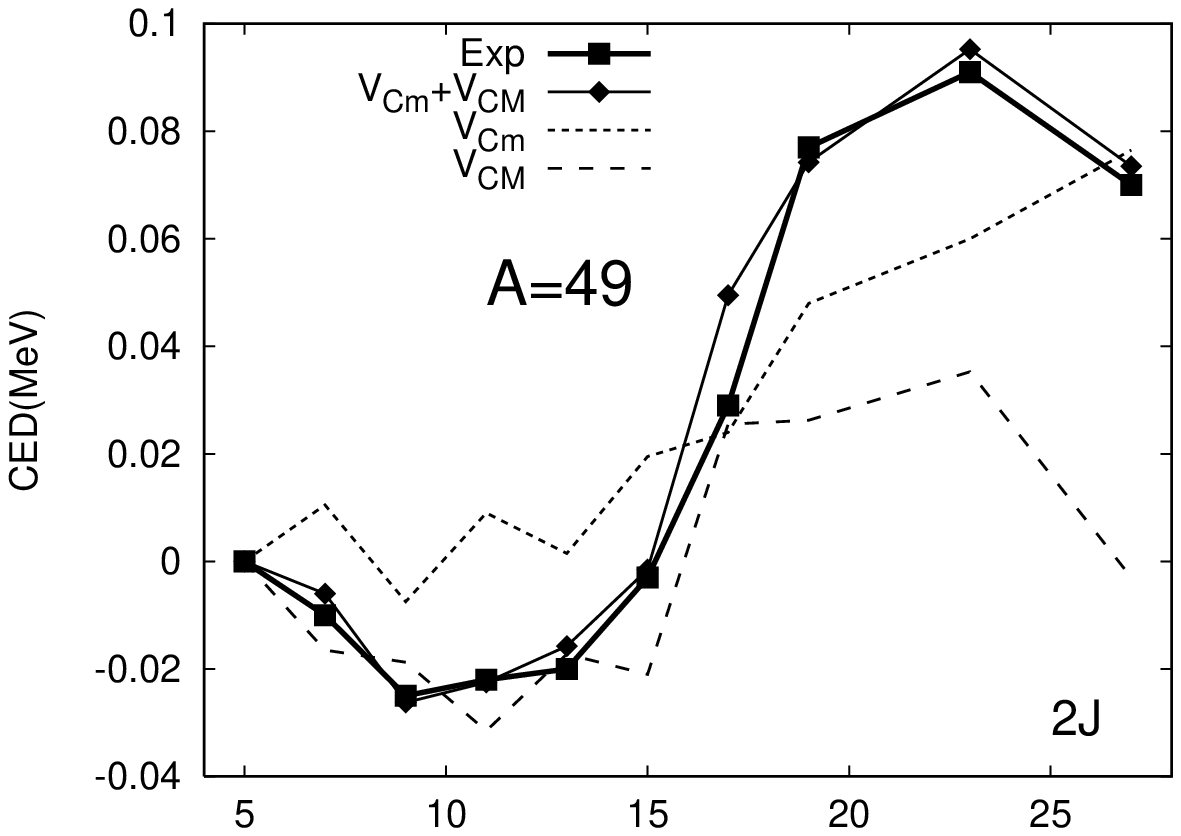,width=7cm}
      \epsfig{file=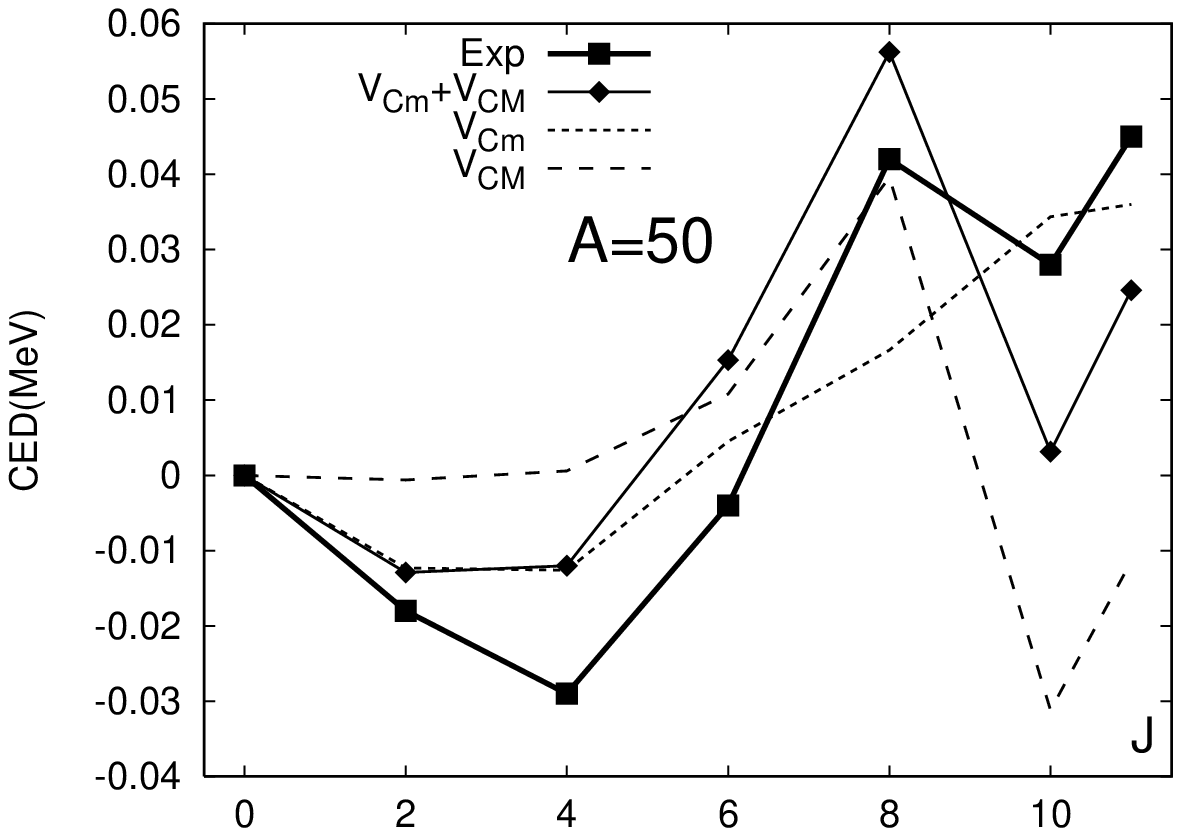,width=7cm}
      \epsfig{file=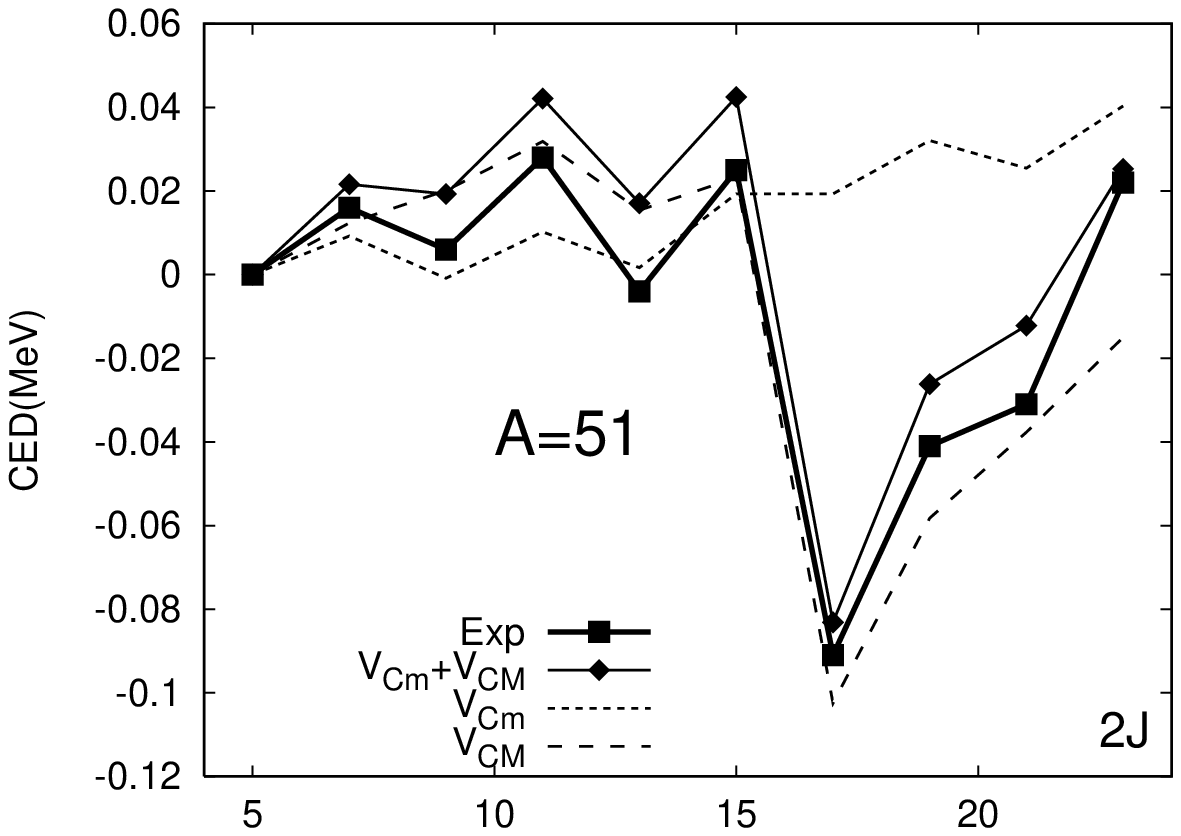,width=7cm}
   \caption {Experimental~\protect\cite{49,47-49,51,50} and
     calculated CED for the pairs $^{47}$Cr-$^{47}$V,
     $^{49}$Mn-$^{49}$Cr, $^{50}$Fe-$^{50}$Cr, and
     $^{51}$Fe-$^{51}$Mn.}
      \label{fig:tutti1}
    \end{center}
\end{figure}
\vspace{-.3cm}
\noindent
Let us write
\begin{eqnarray}
\label{eq:cedj2}
\text{CED}_J&\equiv&V_{Cm}+V_{CM}\\
&=&a_m\langle m_{p_{3/2}}\rangle_J+b_M\, \Delta\langle
V_{Cf_{7/2}}^{eff}\rangle_J+a_M\langle m_{p_{3/2}}\rangle_J
\nonumber
\end{eqnarray}
For $A\approx 50$, $a_m$---estimated at $A=41$---must be reduced by a
factor $(41/50)^{2/3}\approx 0.88$ to account for the $R_0^2$
denominator (after Eq.~(\ref{eq:cedj1})). Therefore, we set
$a_m=0.15$.  Now choose $b_M=b/2$, $a_M=1.8\, a$. Eq.~(\ref{eq:cedj2})
with these parameters (Tab~\ref{tab:param}) yields CED that in
Fig.~\ref{fig:tutti1} are seen too agree well, even very well, with
experiment. The mild exception is $A=50$, discussed in detail in
Ref.~\cite{50}, which contains a heuristic introduction to our CED
results .

The monopole $V_{Cm}$ and multipole $V_{CM}$ contributions, shown
separately in Fig.\ref{fig:tutti1}, indicate that the latter
reproduces only roughly the experimental patterns: The addition of
$V_{Cm}$ is indispensable to bring quantitative agreement. It is
especially worth noting that the strong signature effect in the $A=49$
band is erased in the CED by the out-of-phase $V_{Cm}$ and $V_{CM}$.
Conversely, the signature staggering is enhanced in $A=51$.

The monopole contribution provides valuable information about the
evolution of yrast radii. As a consequence, the use---and even the
validity---of the schematic multipole term (the ``invention'') must be
assessed by the focus it brings to the monopole one, which {\em must}
be present in a form very close to that in Eq.~(\ref{eq:cedj1}).

\vspace{.2cm}
 
To conclude: once the NS anomaly is resolved, the Coulomb field
fulfills the---long held---hope of providing information about radii
not directly measured. The $L^2$ terms offers intriguing prospects. A
complete analysis of $V_{Cm}$, including non-diagonal contributions,
is in order to estimate isospin impurities. The renormalization of
$V_{CM}$ remains an open problem.

\vspace{.4cm}
This work owes much to a stay of AZ at the UAM, made possible by a
scholarship of the BBVA foundation.

\end{multicols}
\end{document}